\documentclass[aps,prd,showkeys,showpacs,amssymb,cite,
amsfonts,epsf,preprintnumbers,nofootinbib,superscriptaddress]{revtex4}

\usepackage[dvips]{graphicx}
\usepackage{bm,latexsym,amsmath,amssymb,amsfonts,color}

\newcommand{\be}{\begin{equation}}
\newcommand{\ee}{\end{equation}}
\newcommand{\bear}{\begin{eqnarray}}
\newcommand{\eear}{\end{eqnarray}}
\newcommand{\ba}{\begin{array}}
\newcommand{\ea}{\end{array}}

\begin{document}

\title{Fluid Black Holes with Electric Field}

\author{Inyong Cho}
\email{iycho@seoultech.ac.kr}
\affiliation{School of Liberal Arts,
Seoul National University of Science and Technology, Seoul 01811, Korea}

\begin{abstract}
We investigate the gravitational field of static perfect-fluid
in the presence of electric field.
We adopt the equation of state $p(r)=-\rho(r)/3$ for the fluid
in order to consider the closed ($S_3$) or the open ($H_3$)
background spatial topology.
Depending on the scales of the mass, spatial-curvature and charge
parameters ($K$, $R_0$, $Q$),
there are several types of solutions
in $S_3$ and $H_3$ classes.
Out of them, the most interesting solution is
the Reisner-Norstr\"om type of black hole.
Due to the electric field, there are two horizons in the geometry.
There exists a curvature singularity inside the inner horizon as usual.
In addition, there exists a naked singularity at the antipodal point in $S_3$
outside the outer horizon due to the fluid.
Both of the singularities can be accessed only by radial null rays.
\end{abstract}
\pacs{04.20.Jb,04.70.Bw}
\keywords{black hole, perfect fluid, electric field}
\maketitle

\section{Introduction}
The spatial topology of the Universe is
one of the unresolved problems in cosmology.
From the recent cosmic microwave background radiation data,
the density fraction of the curvature is estimated as
$\Omega_k = 0.000 \pm 0.005$ ($95\%$, {\it Planck} TT+lowP+lensing+BAO)
\cite{Ade:2015xua}.
Because of the observational error,
it is not possible to determine the spatial topology
from the data at the current stage.
Some other efforts have been made in the inflation models in the closed/open universe
\cite{Ellis:2002we,Ellis:2003qz,Labrana:2013oca,Bucher:1994gb,White:2014aua}.
The investigation of primordial density perturbation shows that
the peculiar predictions of those models are beyond the resolution of 
the current observational data.
Therefore, one needs to consider other ways
in order to catch an idea of the background spatial topology,
for example, the investigation of the gravitating localized objects
in different topologies.

The pure closed/open ($S_3/H_3$) spatial topology is achieved by a constant matter field
with the equation of state,
\begin{align}\label{eom0}
p=-\frac{1}{3}\rho = {\rm constant},
\end{align}
where $\rho>0$ for $S_3$ and $\rho<0$ for $H_3$.
The resulting metric is well known as
\begin{align}\label{metric0}
ds^2 = \mp dt^2 +\frac{dr^2}{1-kr^2/R_0^2} +r^2d\Omega_2^2,
\end{align}
where $k=+1/-1$ represents $S_3/H_3$,
and $\rho =\pm 3/(8\pi R_0^2)$.
For $S_3$, the ranges of the radial coordinate,
$0 \leq r \leq R_0$ and $r \geq R_0$, are considered separately.
(We shall call the former $S_3$-I and the latter $S_3$-II.)
For $S_3$-II, we take $g_{00}=+1$ to consider only one time coordinate.

The metric \eqref{metric0} is the only solution to the Einstein's equation
with the matter of Eq.~\eqref{eom0}.
There is no additional mass term unlike in vacuum
which admits  the flat Minkowski space
as the massless limit of the Schwarzschild spacetime.
In order to achieve a nontrivial structure such as a black hole in $S_3/H_3$,
other type of matter than Eq.~\eqref{eom0} needs to be introduced.
Then, the $S_3/H_3$ nature will be exposed only at some place of space
while a nontrivial geometry is formed elsewhere.

For the nontrivial geometrical structure
that admits the inherent $S_3$/$H_3$ topology,
the static fluid configuration with the equation of state $p(r) = -\rho(r)/3$
was recently studied in Ref.~\cite{Cho:2016kpf}.
It was found that
there are a black-hole solution ($S_3$-I, $S_3$-II, $H_3$),
a nonstatic cosmological solution ($S_3$-II, $H_3$),
and a singular static solution ($H_3$).
The nontrivial geometries of these three types of solutions
are sourced by fluid.
At some region of space,
the signature of the $S_3/H_3$ topology appears
(near the equator for $S_3$-I, near the center for $S_3$-II,
and at the asymptotic region for $H_3$).
In this sense, we interpret the nontrivial geometrical configuration
as a gravitating object formed in the $S_3/H_3$ background spatial topology.
This object can be considered as a large fluid object
which is produced in a global universe,
or a local compact object which is produced in a local $S_3$/$H_3$ space.

In this paper, we consider the same static fluid in Ref.~\cite{Cho:2016kpf}
with the electric field in spherical symmetry.
If there is only the electric field,
the spacetime is described by the Reisner-Norstr\"om solution.
If we add the constant matter of Eq.~\eqref{eom0} to the electric field,
there is no consistent static solution to the Einstein's equation.
Therefore, as in Ref.~\cite{Cho:2016kpf} we consider the fluid of $p(r) = -\rho(r)/3$.
The mixture of electric field and fluid form the geometry, and we expect
that the $S_3/H_3$ topology due to fluid unveils at some region of space.
When the electric field is turned off,
the system reduces to the fluid-only case investigated in Ref.~\cite{Cho:2016kpf}.
There are some other works on the gravitating solutions for static fluids
(see e.g.,  Refs.~\cite{Bekenstein:1971ej,Sorkin:1981wd,Pesci:2006sb,Semiz:2008ny,Lake:2002bq,Bronnikov:2008ia,Cho:2017nhx}).

This paper consists as following.
In Sec. II, we introduce the model and field equations.
In Sec. III, we classify the solutions and discuss the spacetime structure.
In Sec. IV, we discuss the geodesic motions.
In Sec. V, we study the stability of the solutions.
In Sec. VI, we conclude.

\section{Model and field equations}
We consider the electric field and the perfect fluid in static state.
The static metric ansatz for spherical symmetry is given by
\begin{align}\label{metricfgr}
ds^2 = -f(r)dt^2 +g(r)dr^2 +r^2d\Omega_2^2.
\end{align}
The energy-momentum tensor for the fluid is given by
\begin{align}\label{emF}
T^\mu_\nu = {\rm diag}[-\rho(r), p(r),p(r),p(r)],
\end{align}
and we consider the equation of state which meets
the $S_3/H_3$ boundary condition,
\begin{align}\label{eos}
p(r)=-\frac{1}{3}\rho(r) .
\end{align}
The field-strength tensor for the electric field is given by
\begin{align}\label{Fmunu}
{\cal F}_{\mu\nu} = \partial_\mu A_\nu - \partial_\nu A_\mu .
\end{align}
We consider the static electric field only,
then the vector potential is given by
\begin{align}
A_\mu = [A_0(r),0,0,0].
\end{align}
Then the nonvanishing components of ${\cal F}_{\mu\nu}$ in Eq.~\eqref{Fmunu} are
\begin{align}\label{Ftr}
{\cal F}_{01} = -{\cal F}_{10} = E(r) = [f(r)A_0(r)]',
\end{align}
where $E(r)$ is the electric field, 
and the prime denotes the derivative with respect to $r$.
The energy-momentum tensor for the electric field is given by
\begin{align}\label{emE}
{\cal T}^\mu_\nu
= {\cal F}^{\mu\alpha}{\cal F}_{\nu\alpha}
-\frac{1}{4}\delta^\mu_\nu {\cal F}_{\alpha\beta}{\cal F}^{\alpha\beta}
= \frac{E^2(r)}{2f(r)g(r)} {\rm diag}(-1,-1,1,1).
\end{align}
With the metric \eqref{metricfgr} and the energy-momentum tensors
\eqref{emF} and \eqref{emE},
the nonvanishing components of the Einstein's equation,
$G^\mu_\nu = 8\pi(T^\mu_\nu +{\cal T}^\mu_\nu$),
are
\begin{align}
G^0_0 &= -\frac1{r^2} + \frac{1}{r^2 g} - \frac{g'}{r g^2}
= - 8\pi \left[ \rho(r) +\frac{E^2(r)}{2f(r)g(r)} \right] ,\label{G00} \\
G^1_1 &= -\frac1{r^2} + \frac{1}{r^2 g}+ \frac{f'}{rfg}
= 8\pi \left[ p(r) -\frac{E^2(r)}{2f(r)g(r)} \right] , \label{G11} \\
G^2_2 &= G^3_3 = \frac{f'}{2rfg} - \frac{f'^2}{4f^2 g} - \frac{g'}{2rg^2} -
	\frac{f' g'}{4f g^2} + \frac{f''}{2fg}
= 8\pi \left[ p(r) +\frac{E^2(r)}{2f(r)g(r)} \right] . \label{G22}
\end{align}
Since the fluid and the electric field are minimally coupled
only thorough gravity,
the conservation of the energy-momentum tensor is satisfied individually,
$\nabla_\mu T^{\mu\nu} = 0$ and $\nabla_\mu {\cal T}^{\mu\nu} = 0$,
which provide the field equations,
\begin{align}\label{TEeqn}
\rho' +\frac{f'}{f}\rho =0,
\qquad
\frac{3E^2}{fg} \left( \frac{E'}{E} -\frac{f'}{2f} -\frac{g'}{2g} +\frac{2}{r} \right) =0.
\end{align}
These field equations give solutions for fluid and electric field
in terms of the gravitational field,
\begin{align}\label{solrhoE}
\rho(r) = {\rm constant} \times f(r),
\qquad
E(r) = {\rm constant} \times \frac{\sqrt{f(r)g(r)}}{r^2}.
\end{align}

\section{Classification of solutions}
With the solutions in Eq.~\eqref{solrhoE} and the equation of state \eqref{eos},
the Einstein equations \eqref{G00}-\eqref{G22} are solved,
\begin{align}
\rho(r) &= -\frac{3}{8\pi\alpha} \left\{ 1\mp \frac{2\alpha |\beta|}{r}
\left[ \beta(r^2+\alpha) \right]^{1/2}
+\frac{Q^2}{3}\left( \frac{1}{\alpha} +\frac{1}{2r^2} \right) \right\}, \label{rho}\\
f(r) &= \frac{\rho(r)}{\rho_c}, \qquad
g^{-1}(r) = -\frac{8\pi}{3} (r^2+\alpha) \rho(r), \label{g} \\
E(r) &= \frac{Q}{3r^2\left[ \beta(r^2+\alpha) \right]^{1/2}}, \label{E}
\end{align}
where, $Q$ is the electric charge, $\alpha$ and $\beta$ are integration constants,
and $\rho_c = -9\beta/(8\pi)$.
The above solutions reduce to those of the fluid-only solutions in Ref.~\cite{Cho:2016kpf}
when $Q=0$, and to the Reisner-Norstr\"om (RN) solution
when $\alpha \to\infty$ and $\beta \to 0$ with $\alpha\beta = {\rm finite} = M^{2/3}$.

In order to catch the idea of the spatial topology,
we transform the radial coordinate $r$ to $\chi$,
and use the metric
\begin{align}\label{metric2}
ds^2 = -f(\chi)dt^2 +g(\chi)d\chi^2 + R_0^2b^2(\chi)d\Omega_2^2,
\end{align}
where $b(\chi)$ is introduced in the subsections below.
We introduced a new parameter $R_0\equiv \sqrt{|\alpha|}$
which is related with the curvature.
In addition,
we introduce another parameter $K \equiv 2R_0^2|\beta|^{3/2}$
interpreted as a mass parameter
analogous to the fluid-only black hole investigated Ref.~\cite{Cho:2016kpf}.
Depending on the signatures of $\alpha$ and $\beta$,
the solutions are classified into three categories.
Two of them meet the $S_3$ boundary condition,
and the other does the $H_3$ condition.
The classes are summarized in Table I.
When both of the parameter $K$ and the charge $Q$ are turned off,
the metric reduces to that of the pure $S_3$/$H_3$ in Eq.~\eqref{metric0}

\begin{table}
\begin{tabular}{|l||c|c|c|}
  \hline
                \quad Class & $\rho(\chi)$ & $f(\chi)$ & $g(\chi)$ \\ \hline\hline
  \quad $S_3$-I $\quad$
                & $\qquad \frac{3}{8\pi R_0^2} \left[ 1- K \cot\chi -\frac{Q^2}{6R_0^2} (1-\cot^2\chi) \right] \qquad$
                & $\qquad \frac{\rho(\chi)}{\rho_c}, \quad (\rho_c>0) \qquad$
                & $\qquad \frac{3}{8\pi \rho(\chi)} \qquad$ \\ \hline
  \quad $S_3$-II
                & $\frac{3}{8\pi R_0^2} \left[ 1 \mp K \tanh\chi -\frac{Q^2}{6R_0^2} (1+\tanh^2\chi) \right]$
                & $\frac{\rho(\chi)}{\rho_c}, \quad (\rho_c<0)$
                & $-\frac{3}{8\pi \rho(\chi)}$ \\ \hline
  \quad $H_3$
                & $-\frac{3}{8\pi R_0^2} \left[ 1 \mp K \coth\chi +\frac{Q^2}{6R_0^2} (1+\coth^2\chi) \right]$
                & $\frac{\rho(\chi)}{\rho_c}, \quad (\rho_c<0)$
                & $-\frac{3}{8\pi \rho(\chi)}$ \\
  \hline
\end{tabular}
\caption{Classification of solutions.
The signature of $\rho_c$ is chosen so that $f(\chi)g(\chi)>0$.
}
\end{table}

\subsection{$S_3$-I}
This is the case of $\alpha<0$ and $\beta<0$.
The transformation is performed by
\begin{equation}\label{rS3I}
r = R_0b(\chi) = R_0\sin\chi
\quad (0\leq \chi \leq \pi,\; 0\leq r \leq R_0).
\end{equation}
Note that for a given value of $r$, $\chi$ is double valued.
The metric becomes
\begin{equation}\label{metricS3I}
ds^2 = -\frac{3}{8\pi R_0^2\rho_c} \left[ 1- K \cot\chi
-\frac{Q^2}{6R_0^2} (1-\cot^2\chi) \right] dt^2
+\frac{R_0^2}{1- K \cot\chi
-(Q^2/6R_0^2) (1-\cot^2\chi)} d\chi^2
+R_0^2\sin^2\chi d\Omega_2^2.
\end{equation}
Here, $\rho_c>0$.
This solution states that the fluid with the electric field strength in Eq.~\eqref{Ftr}
closes the space in a finite region $ 0\leq r \leq R_0$.
We believe that the fluid is responsible for this closure
since the same phenomenon occurs even in the fluid-only case in Ref.~\cite{Cho:2016kpf}.
Both of the Ricci scalar and the Kretschmann scalar diverge
at $\chi=0$ and $\pi$, i.e.,
there exist curvature singularities at both poles.
For the pure fluid case ($Q=0$) investigated in Ref.~\cite{Cho:2016kpf},
the background $S_3$ topology is exposed at the boundary around the equator
($\chi \approx\pi/2$, i.e., $r\approx R_0$),
\begin{align}
ds_3^2 \approx R_0^2d\chi^2 +R_0^2\sin^2\chi d\Omega_2^2.
\end{align}
With the electric field, however, there is a charge correction,
\begin{align}
ds_3^2 \approx \frac{R_0^2}{1-Q^2/(6R_0^2)}d\chi^2 +R_0^2\sin^2\chi d\Omega_2^2.
\end{align}
The location of the horizon is found from $g_{\chi\chi}^{-1}=0$,
\begin{equation}
\chi_h = \chi_\pm \equiv \cot^{-1} \left( \frac{3KR_0^2 \mp \sqrt{J_1}}{Q^2} \right),
\quad\mbox{where}\quad
J_1=9K^2R_0^4 -6Q^2R_0^2+Q^4.
\end{equation}
Depending on the existence of the horizon, there are two types of solutions.
(See Fig. 1 for the graphical view of the metric function.)

(i) {\bf RN black-hole type solution}:
If $J_1>0$, there exist two horizons at $\chi_h=\chi_\pm$,
which coalesce when $J_1=0$.
This solution mimics the Reisner-Nordstr\"om geometry of the charge black hole.
The spacetime is regular at $\chi<\chi_-$ and $\chi>\chi_+$.
The singularity at the north pole ($\chi=0$) is inside the inner horizon,
and is not accessible by the timelike observers as in the RN black hole.
The singularity at the south pole ($\chi=\pi$) is naked,
but is not accessible either by the timelike observers as in the fluid black hole
investigated in Ref.~\cite{Cho:2016kpf}.
The geodesics are studied in the next section.

(ii) {\bf Naked singular solution}:
If $J_1<0$, there is no horizon.
Both singularities are naked, but neither of them are accessible.

\begin{figure*}[btph]
\begin{center}
\includegraphics[width=0.3\textwidth]{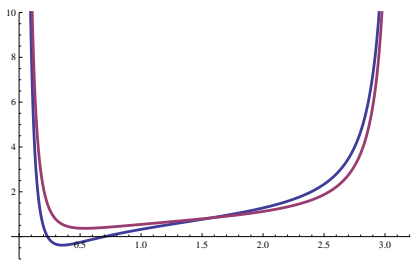}
\end{center}
\caption{Plot of metric function $8\pi R_0^2 \rho(\chi)/3$ for $S_3$-I.
(i) RN black-hole type solution: $K=0.9$, $Q=1$, $R_0=1$.
There are two horizons between which the spacetime is nonstatic.
There exist two curvature singularities. 
Neither of them is accessible except by radial null rays.
(ii) Naked singular solution: $K=5/9$, $Q=1$, $R_0=1$.
}
\end{figure*}

\subsection{$S_3$-II}
This is the case of $\alpha<0$, $\beta>0$.
The transformation is performed by
\begin{equation}\label{rS3II}
r=R_0b(\chi) =R_0\cosh\chi
\quad ( -\infty < \chi < \infty,\; r\geq R_0),
\end{equation}
Again, $\chi$ is double-valued for a given value of $r$.
The metric becomes
\begin{equation}\label{metricS3II}
ds^2 = -\frac{3}{8\pi R_0^2\rho_c} \left[ 1 \ominus\oplus K \tanh\chi
-\frac{Q^2}{6R_0^2} (1+\tanh^2\chi)  \right] dt^2
+\frac{R_0^2}{-\left[ 1 \ominus\oplus K \tanh\chi
-(Q^2/6R_0^2) (1+\tanh^2\chi)  \right]} d\chi^2
+R_0^2\cosh^2\chi d\Omega_2^2.
\end{equation}
Here, $\rho_c<0$.
The fluid curves the space in a flipped way to the $S_3$-I case;
the space is confined in the open region $r\geq R_0$.
The curvature is finite everywhere.
The location of the horizon is
\begin{equation}
\chi_h = \chi_\pm \equiv \tanh^{-1} \left( \frac{\ominus\oplus 3KR_0^2 \pm \sqrt{J_2}}{Q^2} \right),
\quad\mbox{where}\quad
J_2=9K^2R_0^4 +6Q^2R_0^2 -Q^4.
\end{equation}
(The $\pm$ roots are valid for both $\ominus$ and $\oplus$.)
%
There are four types of solutions. (See Fig. 2.)
Two of them are black-hole type solutions
(Schwarzschild and Reisner-Nordstr\"om types)
without a singularity,
and the others are regular and nonstatic solutions.

\vspace{12pt}
Let us consider the  $\ominus$ solution.
\vspace{12pt}

If $J_2>0$, there are three types of solutions.

(i) {\bf RN black-hole type solution}:
For $Q^2 \geq 3(1+K)R_0^2$,
there are two horizons at $\chi_\pm$
and this is the RN black-hole type.

(ii) {\bf Schwarzschild black-hole type solution}:
For $3(1-K)R_0^2 < Q^2 < 3(1+K)R_0^2$,
there exists only one horizon.
Inside the horizon (the trapped region), $f(\chi), g(\chi) <0$ and $\rho>0$.
The spacetime is nonstatic in the trapped region, and static outside.
The structure is similar to that of the Schwarzschild black hole.

(iii) {\bf Nonstatic solution}:
For $Q^2 \leq 3(1-K)R_0^2$, there is no horizon and the spacetime is nonstatic everywhere.
This type of solution is special for $S_3$-II.
This is analogous to the solution in  Eq.~\eqref{metric0}
describing the region $r\geq R_0$
in which the roles of the temporal and the radial coordinates are exchanged.

\vspace{12pt}
If $J_2<0$, there is one type of solution.

(iv) {\bf Regular solution}:
The spacetime is regular everywhere while $\rho<0$.

\vspace{12pt}
For the $\oplus$ solution,
the situation is the same with the $\ominus$ solution
with $\chi \to -\chi$.
Therefore, out of four types (i)-(iv),
the only change is in (ii).
Now, the region of $\chi < \chi_h$ is static,
and the region of  $\chi > \chi_h$ is nonstatic.

\begin{figure*}[btph]
\begin{center}
\includegraphics[width=0.3\textwidth]{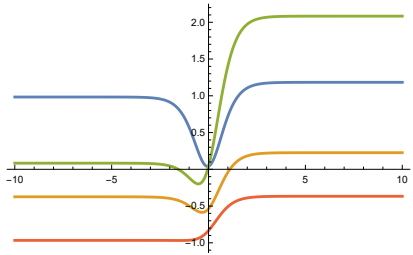}
\end{center}
\caption{Plot of metric function $-8\pi R_0^2 \rho(\chi)/3$ for $S_3$-II.
(i) RN black-hole type solution: $K=1$, $Q=1$, $R_0=0.4$.
(ii) Schwarzschild black-hole type solution: $K=0.3$, $Q=1$, $R_0=0.6$.
(iii) Nonstatic solution: $K=0.1$, $Q=1$, $R_0=0.4$.
(iv) Regular solution: $K=0.3$, $Q=1$, $R_0=1$.
}
\end{figure*}

\subsection{$H_3$}
This is the case of $\alpha>0$, $\beta>0$.
The transformation is performed by
\begin{equation}\label{rH3}
r= R_0b(\chi) =R_0\sinh\chi \quad (\chi \geq 0,\; r \geq 0),
\end{equation}
and the metric becomes
\begin{equation}\label{metricH3}
ds^2 = -\frac{3}{8\pi R_0^2(-\rho_c)} \left[ 1 \ominus\oplus K \coth\chi
+\frac{Q^2}{6R_0^2} (1+\coth^2\chi)  \right] dt^2
+\frac{R_0^2}{1 \ominus\oplus K \coth\chi +(Q^2/6R_0^2) (1+\coth^2\chi)} d\chi^2
+R_0^2\sinh^2\chi d\Omega_2^2.
\end{equation}
Here, $\rho_c<0$.
The curvature diverges at $\chi=0$.
The location of the horizon is
\begin{equation}
\chi_h = \chi_\pm \equiv \coth^{-1} \left( \frac{\oplus\ominus 3KR_0^2 \mp \sqrt{J_3}}{Q^2} \right),
\quad\mbox{where}\quad
J_3=9K^2R_0^4 -6Q^2R_0^2 -Q^4.
\end{equation}
The solutions are classified as below. (See Fig. 3.)

\vspace{12pt}
For the $\ominus$ solution in Eq.~\eqref{metricH3},
there are three types of solutions for $J_3>0$.
\vspace{12pt}

(i) {\bf RN black-hole type solution}:
For $3(K-1)R_0^2 < Q^2 < 3KR_0^2$, there are two horizons at $\chi_\pm$
and this is the RN black-hole type.

(ii) {\bf dS-type solution}:
For $Q^2 \leq 3(K-1)R_0^2 $, there is only one horizon at $\chi_+$.
The spacetime is static inside the horizon, and nonstatic outside.
This is a de Sitter-like solution.
This solution is achieved when the electric charge $Q$ is small.
When $Q=0$, this corresponds to the cosmological solution
of the fluid-only case in Ref.~\cite{Cho:2016kpf}
for which the spacetime is nonstatic everywhere.
It was interpreted as a universe expanding from an initial singularity.
For the present case, however, 
the horizon is formed due to the electric field
inside which the spacetime is static.

(iii) {\bf Naked singular solution}:
For $Q^2 \geq 3KR_0^2$, the solution is static everywhere,
but with a singularity at the center.

\vspace{12pt}
For the $\oplus$ solution in Eq.~\eqref{metricH3}, or for $J_3<0$,
there is no horizon, and the solution is singular static like (iii).

\begin{figure*}[btph]
\begin{center}
\includegraphics[width=0.3\textwidth]{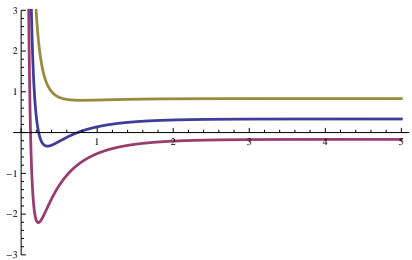}
\end{center}
\caption{Plot of metric function $-8\pi R_0^2 \rho(\chi)/3$ for $H_3$.
(i) RN black-hole type solution (blue): $K=1$, $Q=1$, $R_0=1$.
(ii) dS-type solution (red): $K=1.5$, $Q=1$, $R_0=1$.
(iii) Naked singular solution: $K=0.5$, $Q=1$, $R_0=1$.
}
\end{figure*}

\subsection{Gauss' Law}
Let us discuss the Gauss' law in the $\chi$ coordinate.
The field-strength tensor ${\cal F}_{\mu\nu}$ in Eq.~\eqref{Fmunu}
in the $r$ coordinate with the components \eqref{Ftr}
is transformed to ${\cal F}'_{\mu\nu}$ in the $\chi$ coordinate
with the nonzero components,
\begin{align}\label{Ftchi}
{\cal F}'_{t\chi} = -{\cal F}'_{\chi t} = E(\chi)
= \frac{Q}{3|\beta|^{1/2}R_0^2b^2(\chi)} .
\end{align}
The electric flux is then
\begin{align}
\Phi_E= \oint E\sqrt{g^{(2)}}d^2x
= \iint \frac{Q}{3|\beta|^{1/2}R_0^2b^2(\chi)}
\times R_0^2b^2(\chi) \sin\theta d\theta d\phi
= \frac{4\pi Q}{3|\beta|^{1/2}}
= \frac{4\pi Q}{\sqrt{8\pi |\rho_c|}},
\end{align}
where we used the relation $\rho_c = -9\beta/(8\pi)$.
Compared with the Gauss' law in flat space,
there is a correction due to fluid by the factor $\sqrt{8\pi |\rho_c|}$.

\subsection{Mass}
In this section, let us discuss the mass of the black-hole solutions.
For the fluid-only case in Ref.~\cite{Cho:2016kpf},
it was investigated that the horizon structure of the fluid black hole
is similar to that of the Schwarzschild black hole.
The parameters are related with the Schwarzschild mass $M$ as
\begin{align}\label{KM}
K = \left( \frac{R_0^2}{4M^2} -1 \right)^{-1/2},
\quad
\left( -\frac{R_0^2}{4M^2} +1 \right)^{-1/2},
\quad
\left( \frac{R_0^2}{4M^2} +1 \right)^{-1/2},
\end{align}
for the type $S_3$-I, $S_3$-II, and $H_3$, respectively.
For the $S_3$-I type, 
there is an upper limit in the mass, $M \to R_0/2$ as $ K \to \infty$.
In this limit, the horizon approaches the equator of $S_3$, $\chi_h = \cot^{-1}(1/K) \to \pi/2$.

Other than the Schwarzschild mass, it is interesting to consider the Misner-Sharp mass ${\cal M}$
which can be used for black-hole thermodynamics \cite{Misner:1964je}.
We evaluate ${\cal M}$ in this work.
When the metric is given by
\begin{align}\label{metricMS}
ds^2 = h_{ab} dx^adx^b +r^2(x)d\Omega_2^2,
\end{align}
where $a,b=0,1$, 
the Misner-Sharp mass is defined as
\begin{align}\label{MS}
{\cal M} = \frac{1}{2} (1-h^{ab} \partial_a r \partial_b r).
\end{align}
In the $\chi$ coordinate, we have $r=R_0b(\chi)$
and Eq.~\eqref{MS} becomes
\begin{align}
{\cal M}(\chi) = -\frac{4\pi R_0^3}{3s} \rho(\chi) b(\chi)  [b'(\chi)]^2 +\frac{R_0}{2}b(\chi),
\end{align} 
where $s$ is the signature of $\rho_c$
($s=+1$ for $S_3$-I, and $s=-1$ for the others).
The mass depends on the radial coordinate $\chi$.

For the fluid-only case ($Q=0$), the mass is
still $\chi$ dependent,
while one has ${\cal M}_{\rm Sch} = M$ for the ordinary Schwarzschild black hole.
For the fluid black-hole solutions,  one can show with the aid of Eq.~\eqref{KM} that
the Misner-Sharp mass evaluated on the horizon 
coincides with the Schwarzschild mass,
${\cal M}(\chi_h) = M$.
This indicates that the horizon structure of the fluid black hole 
is the same with that of the Schwarzschild black hole.

For the ordinary RN black hole, the Misner-Sharp mass is 
given by ${\cal M}_{\rm RN} =M-Q^2/(2r) = M-Q^2/[2R_0b(\chi)]$.
For the RN black-hole type solutions obtained in this work ($Q\neq 0$),
keeping the mass relation of $K$ in Eq.~\eqref{KM},
the Misner-Sharp mass evaluated on the horizons does not
coincide with that of the ordinary RN black hole,
${\cal M}(\chi_\pm) \neq  {\cal M}_{\rm RN}(\chi_\pm)$.

Although the horizon structure of the fluid black hole ($Q=0$)
is the same with that of the ordinary one,
the thermodynamics must be very different because the off-horizon structure
is very different. 
We shall study the thermodynamics using the Misner-Sharp mass
in a separate work including the charged case.

\section{Geodesics}
In this section, we discuss the geodesics of the solutions.
We focus mainly on the black-hole solutions.
For simplicity, we define a function,
\begin{align}\label{Fchi}
F(\chi) \equiv  \frac{8\pi R_0^2}{3s}\rho(\chi).
\end{align}
The geodesic equations become
\begin{align}
&\mbox{$t$-eq. : }\quad
\frac{1}{F(\chi)} \frac{d}{d\lambda} \left[ F(\chi) \frac{dt}{d\lambda}\right]
=0 ,\label{S3teq}\\
&\mbox{$\phi$-eq. : }\quad
\frac{1}{b^2(\chi)} \frac{d}{d\lambda} \left[ b^2(\chi) \frac{d\phi}{d\lambda}\right]
=0.\label{S3phieq}
\end{align}
From Eqs.~\eqref{S3teq} and \eqref{S3phieq},
we denote the conserved quantities 
$E$ (energy) and $L$ (angular momentum) as
\begin{align}
E \equiv F(\chi) \frac{dt}{d\lambda} = {\rm constant},
\qquad
L \equiv b^2(\chi) \frac{d\phi}{d\lambda} = {\rm constant}.
\end{align}
The $\chi$-equation can be derived from the metric as
\begin{align}\label{chieq}
g_{\mu\nu} \frac{dx^\mu}{d\lambda} \frac{dx^\nu}{d\lambda} = - \varepsilon,
\end{align}
where $\varepsilon = 0,1$ for null and timelike geodesics, individually.
On the $\theta=\pi/2$ plane, Eq.~\eqref{chieq} becomes
\begin{align}
\frac{1}{2} \left( \frac{d\chi}{d\lambda} \right)^2 + V(\chi) = \frac{3E^2}{16\pi R_0^4|\rho_c|} \equiv \tilde{E}^2,
\end{align}
where the effective potential is given by
\begin{align}
V(\chi) = \frac{1}{2} F(\chi) \left[ \frac{L^2}{b^2(\chi)} +\frac{\varepsilon}{R_0^2} \right].
\end{align}
We summarize $V(\chi)$ in Table II.
The effective potential $V(\chi)$ of the black-hole type solutions
is plotted in Fig. 4-6.

For the  RN black-hole type solution of $S_3$-I, 
the singularities at both poles are not accessible
except by the radial null geodesic.
For the fluid-only case in Ref.~\cite{Cho:2016kpf},
the one at the north pole inside the horizon was accessible
since the inner geometry was similar to that of the Schwarzschild black hole.
However, for the present case, it is not because 
the inner geometry is similar to that of the charged black hole.
The nonaccessibility to the naked singularity at the south pole
is similar to the fluid-only case.
The geodesic observer starting from the outer static region 
falls into the inner static region passing the intermediate nonstatic region.
Afterwards, the observer bounces back to the nonstatic region and then
enters the outer static region.
This later motion after the bounce proceeds
in the other copy of the spacetime
accompanied in the usual RN geometry.
The geodesic as a whole is an oscillatory orbit
in the infinite tower of the RN spacetime.

For $S_3$-II, the RN black-hole type solution, 
when the energy level ($\tilde E$) is low,
the oscillatory orbit is similar to 
that of $S_3$-I.
When the energy level is increased,
the geodesic observer can reach the inner static region behind the inner horizon.
When the energy level is high enough,
the geodesic observer can escape to the asymptotic infinity in the static region.
The Schwarzschild black-hole type solution has
the similar geodesic structure to that of the usual Schwarzschild black hole.
When the energy level is low,
all the geodesic motions fall into the black hole.
However, $V(\chi)$ approaches a constant value as $\chi \to -\infty$.

For $H_3$, the singularity at the center is not accessible
except by the radial null geodesic,
which is different from the fluid-only case.
Similarly to the $S_3$-I,
it is due the electric charge.
When the energy level is low,
the geodesic motion is oscillatory
as in $S_3$-I.
When the energy level is high,
the geodesic observer can reach the asymptotic infinity.
Another interesting solution is dS-type.
For this solution, the geodesics escape from the static region
crossing the de Sitter-like horizon and reach asymptotic infinity.
This is different from the pure de Sitter space
in which there can be a stable geodesic motion inside the horizon.

\begin{table}
\begin{tabular}{|l||c|c|}
  \hline
                \quad Class & $F(\chi)$ & $V(\chi)$ \\ \hline\hline
  \quad $S_3$-I $\quad$
                & $\qquad 1- K \cot\chi -(Q^2/6R_0^2) (1-\cot^2\chi) \qquad$
                & $\qquad \frac{1}{2} [1- K \cot\chi -(Q^2/6R_0^2) (1-\cot^2\chi)] \left( \frac{L^2}{\sin^2\chi} +\frac{\varepsilon}{R_0^2} \right) \qquad$ \\ \hline
  \quad $S_3$-II
                & $-1 \pm K \tanh\chi +(Q^2/6R_0^2) (1+\tanh^2\chi) $
                & $\frac{1}{2} [-1 \pm K \tanh\chi +(Q^2/6R_0^2) (1+\tanh^2\chi)] \left( \frac{L^2}{\cosh^2\chi} +\frac{\varepsilon}{R_0^2} \right)$ \\              \hline
  \quad $H_3$
                & $1 \mp K \coth\chi +(Q^2/6R_0^2) (1+\coth^2\chi)$
                & $\frac{1}{2} [1 \mp K \coth\chi +(Q^2/6R_0^2) (1+\coth^2\chi)] \left( \frac{L^2}{\sinh^2\chi} +\frac{\varepsilon}{R_0^2} \right)$ \\
  \hline
\end{tabular}
\caption{Effective potential $V(\chi)$}
\end{table}

\begin{figure*}[btph]
\begin{center}
\includegraphics[width=0.3\textwidth]{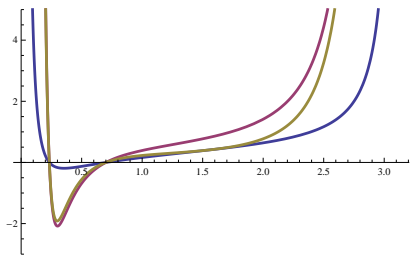}
\end{center}
\caption{Plot of effective potential $V(\chi)$
fo (i) RN black-hole type solution ($K=0.9$, $Q=1$, $R_0=1$) 
for $S_3$-I.
[$L=0$ (blue), $L=1$ (red) for timelike and $L=1$ for null.]
The shape of the potential shows that the two singularities 
are not accessible except by the radial ($L=0$) null geodesic.
The geodesic observers can get into the inner region of the black hole.
Then they bounce to the outer region in the other copy of the spaceitme
as usual in the Reisner-Norstr\"om geometry 
in which there exists an infinite tower of spacetime.
}
\end{figure*}

\begin{figure*}[btph]
\begin{center}
\includegraphics[width=0.3\textwidth]{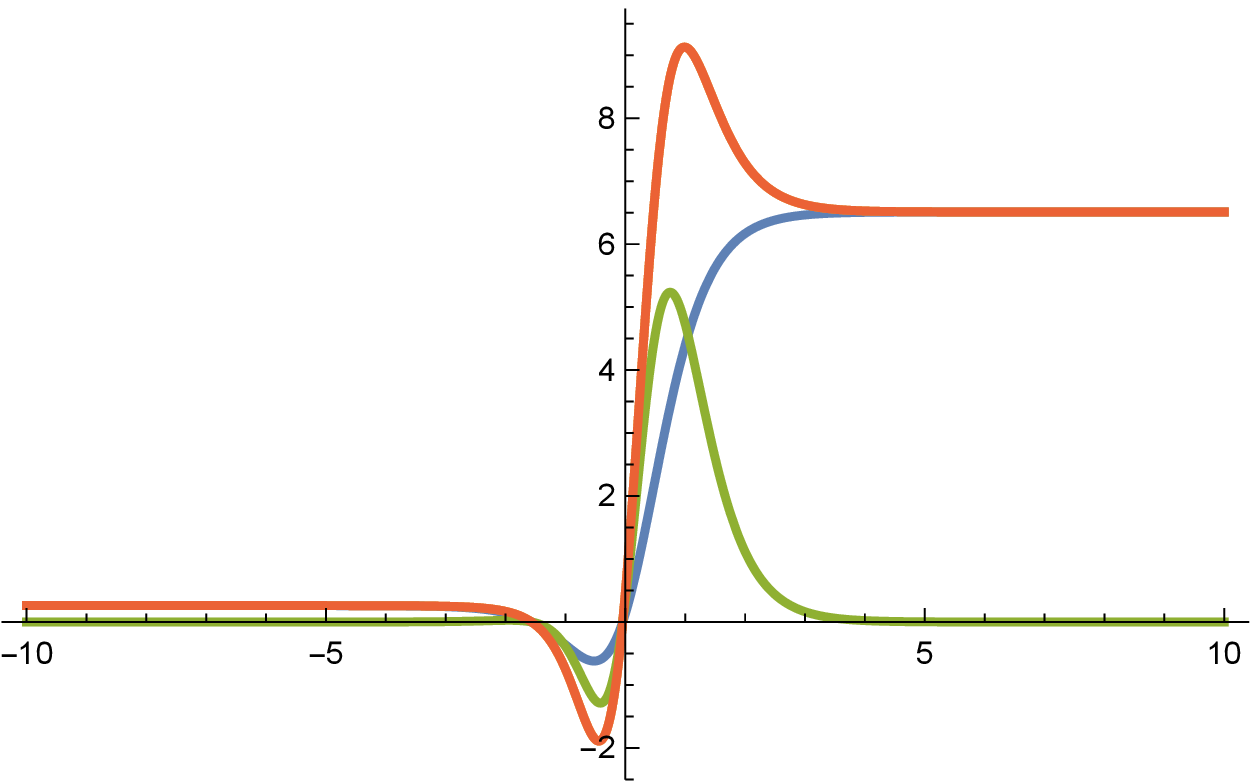}
\includegraphics[width=0.3\textwidth]{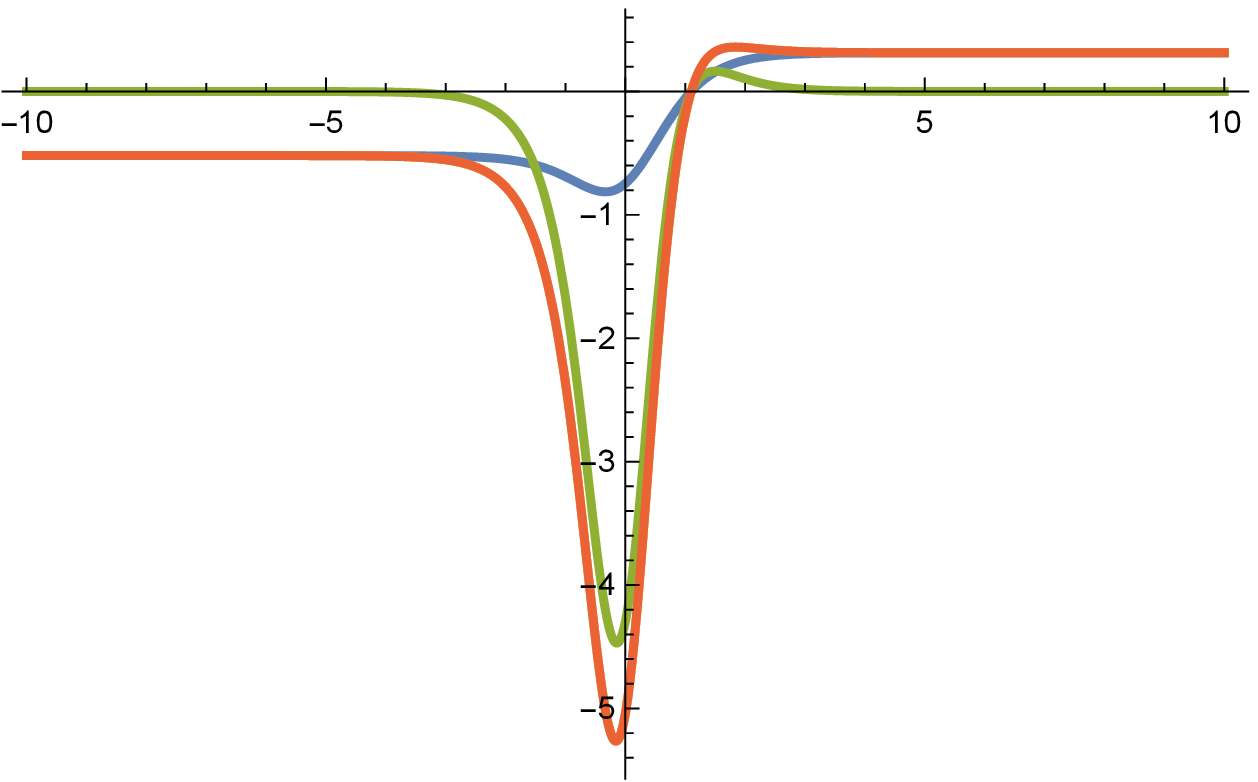}\\
(i) \hspace{2in} (ii)\\
\end{center}
\caption{Plot of effective potential $V(\chi)$ for $S_3$-II.
[$L=0$ (blue), $L=4$ (red) for timelike.]
(i) RN black-hole type solution: $K=1$, $Q=1$, $R_0=0.4$, $L_{\rm null}=4$.
For the low energy level ($\tilde E$), 
the geodesic motion is similar to that of $S_3$-I (i),
which oscillates in the infinite spacetime tower.
For the intermediate energy level,  
the geodesic motion can reach the inner static region behind the inner horizon.
For the high energy level,
the geodesic motion can reach the asymptotic infinity at the outer static region.
(ii) Schwarzschild black-hole type solution: $K=0.3$, $Q=1$, $R_0=0.6$, $L_{\rm null}=4$.
The potential is similar to that of the usual Schwarzschild black hole.
For the low energy level, the geodesic motion
falls into the black hole.
}
\end{figure*}

\begin{figure*}[btph]
\begin{center}
\includegraphics[width=0.3\textwidth]{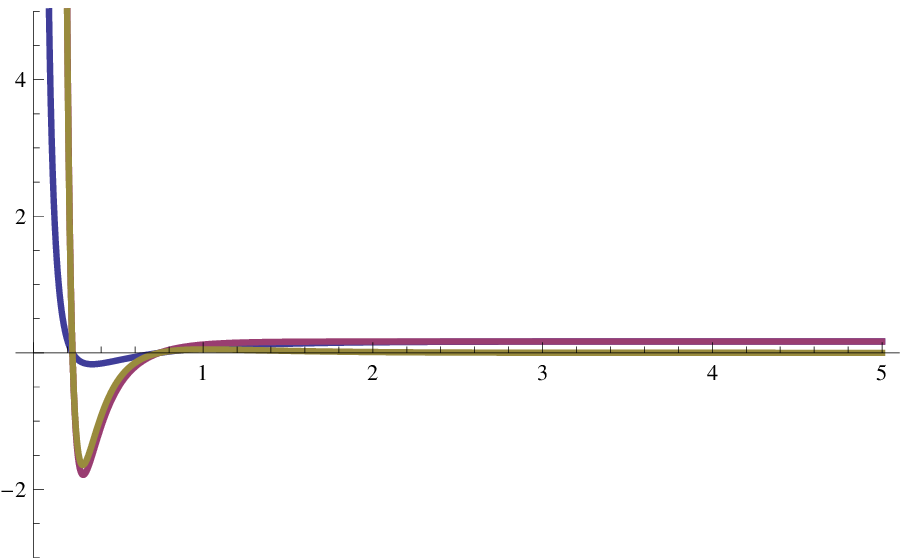}
\includegraphics[width=0.3\textwidth]{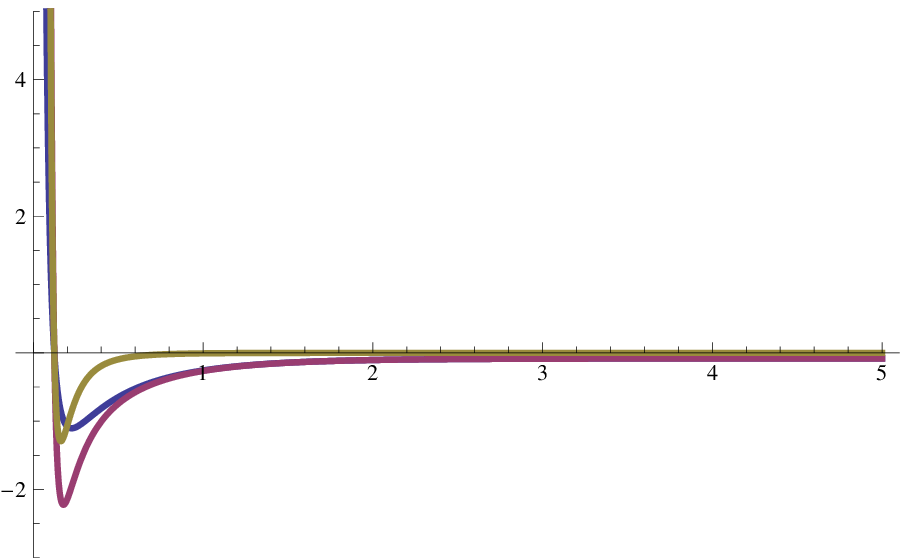}\\
(i) \hspace{2in} (ii)
\end{center}
\caption{Plot of effective potential $V(\chi)$ for $H_3$.
[$L=0$ (blue), $L=1$ (red) for timelike.]
(i) RN black-hole type solution: $K=1$, $Q=1$, $R_0=1$, $L_{\rm null}=1$.
The central singularity is not accessible.
For the high energy level,
the geodesic motion can reach the asymptotic infinity.  
(ii) dS-type solution: $K=1.5$, $Q=1$, $R_0=1$, $L_{\rm null}=0.2$.
There is no stable geodesic motion inside the de Sitter-like horizon.
All the geodesics escape from the static region crossing the horizon.
}
\end{figure*}

\section{Stability}
In this section, we study the stability of the solutions.
We introduce linear spherical scalar perturbations 
with the metric ansatz,
\begin{align}
ds^2 = -f(t,\chi)dt^2 + g(t,\chi) d\chi^2 + R_0^2 b^2(\chi) d\Omega_2^2.
\end{align}
The metric perturbations are introduced as
\begin{align}
f(t,\chi) &= f_0(\chi) + \epsilon f_1(t,\chi), \label{p1}\\
g(t,\chi) &= R_0^2 \big[ g_0(\chi) + \epsilon g_1(t,\chi) \big], \label{p2}
\end{align}
where $\epsilon$ is a small parameter,
and the subscript $0$ stands for the background solutions
obtained in Sec. III.
Using  the function $F(\chi) = 8\pi R_0^2\rho_0(\chi)/3s$ 
defined in Eq.~\eqref{Fchi},
where $\rho_0(\chi)$ is the background solution in Table I,
we have
\begin{align}
f_0(\chi) &= \frac{\rho_0(\chi)}{\rho_c} = \frac{3s}{8\pi R_0^2\rho_c}F(\chi), \\
g_0(\chi) &= \frac{1}{F(\chi)}.
\end{align}
The contravariant form of the energy-momentum tensor for fluid is written as
\begin{align}
T^{\mu\nu} = (\rho +p)u^\mu u^\nu + pg^{\mu\nu},
\end{align}
with the velocity four-vector
\begin{align}
u^\mu = \big[ u^0(t,\chi),u^1(t,\chi),0,0 \big].
\end{align}
For the  fluid at hand, $p=-\rho/3$,
the perturbations for the energy density and the four-velocity are introduced by
\begin{align}
\rho(t,\chi) &= \rho_0(\chi) +\epsilon \rho_1(t,\chi), \label{p3}\\
u^0(t,\chi) &= u_0^0(\chi) +\epsilon u_1^0(t,\chi), \label{p4}\\
u^1(t,\chi) &= u_0^1(\chi) +\epsilon u_1^1(t,\chi). \label{p5}
\end{align}
We have $u_0^1(\chi)=0$ for the comoving background fluid.
From the normalization $u^\mu u_\mu=-1$, we have
$u_0^0(\chi)=1/\sqrt{f_0(\chi)}$ and
$u_1^0(t,\chi)=-f_1u_0^0/(2f_0) = -f_1/(2f_0^{3/2})$.

For the electric field, we introduce the simplest perturbation
along the radial direction only,
by which there is no magnetic field induced by the perturbation,
\begin{align}\label{p6}
{\cal F}'_{t\chi} = -{\cal F}'_{\chi t} = E(t,\chi) = E_0(\chi) + \epsilon E_1(t,\chi),
\end{align}
where $E_0(\chi)$ is given in Eq.~\eqref{Ftchi}.

Now we apply the perturbations \eqref{p1}, \eqref{p2}, and \eqref{p3}-\eqref{p6},
and expand the field equations in the first order of $\epsilon$.
From the $(0,1)$ component of the Einstein's equation, we get
\begin{align}
u_1^1(t,\chi) = -\sqrt{\frac{2\pi R_0^2\rho_c}{3}} \frac{\dot{g_1}b'F}{s^2b\sqrt{F}}.
\end{align}
Therefore, the perturbations of the four-vector, $u_1^0$ and $u_1^1$ in Eqs.~\eqref{p4} and \eqref{p5},
are expressed by the background functions and the metric perturbations.
There are seven equations in total for four perturbations,
$f_1$, $g_1$, $\rho_1$ and $E_1$;
three from Einstein's equation, 
two from $\nabla_\mu T^{\mu\nu} =0$ ,
and two from $\nabla_\mu {\cal T}^{\mu\nu} =0$.
Four of them are independent equations.
After manipulating equations with
\begin{align}
f_1(t,\chi) &= e^{i\omega t} \psi(\chi),\\
g_1(t,\chi) &= e^{i\omega t} \varphi(\chi),
\end{align}
the equation for $\varphi(\chi)$ is decoupled as
\begin{align}\label{PE1}
-F^2\varphi''
-\left[ 3FF' + F^2 \left( 3\frac{b''}{b'} +s\frac{b}{b'} \right) \right] \varphi'
+\left[ \frac{\omega^2}{\sigma} -2FF''
-FF'\left( 4\frac{b''}{b'}-\frac{b'}{b} -s\frac{b}{b'} \right)
-2F^2 \left( \frac{b'''}{b'} -\frac{b'^2}{b^2} +s \frac{bb''}{b'^2} -s \right)
\right] \varphi =0,
\end{align}
where $\sigma \equiv 1/(8\pi R_0^4\rho_c s) = 1/(8\pi R_0^4|\rho_c|) >0$ for all classes.
The coefficients of the above equation depend only on the background functions $F(\chi)$ and $b(\chi)$.

By transforming the radial coordinate and the amplitude function as
\begin{align}
z = \int^\chi_0 \frac{d\chi}{\sqrt{2}F(\chi)}, \qquad
\Phi(z) = N \frac{F(\chi)b'(\chi)}{z} \varphi(\chi),
\end{align}
where $N$ is a normalization constant,
we get the perturbation equation in the nonrelativistic Schr\"odinger-type,
\begin{align}\label{PE2}
\left[ -\frac{1}{2}\frac{d^2}{dz^2} - \frac{1}{z}\frac{d}{dz}
+U(z) \right] \Phi(z) = -\frac{\omega^2}{\sigma} \Phi(z)
=-8\pi R_0^4|\rho_c| \omega^2 \Phi(z) \equiv \Omega \Phi(z).
\end{align}
The potential is given by
\begin{align}
U[z(\chi)] = F^2 \left[ -\frac{F''}{F} +\left( \frac{F'}{F} \right)^2
+\frac{F'}{F} \left( \frac{b''}{b'} +2\frac{b'}{b} +4s \right)
+2 \left( \frac{b''}{b'} \right)^2 +s
\right],
\end{align}
where we used $sb/b' = -b''/b'$, $b'''/b'=-s$, and $b''/b=-s$.
Since there always exists a positive eigenvalue $\Omega$ for any type of potential $U$,
i.e., $\omega^2 <0$,
this system is {\it unconditionally unstable}.

The stability story is very similar to the fluid-only case.
When perturbations are introduced to the static fluid,
the fluid becomes time dependent,
which drives the Universe to undergo the Friedmann expansion.
This type of instability does not necessarily mean 
that the black-hole structure is destroyed.
Instead, the instability indicates that the background universe
undergoes expansion while the black-hole structure sustains.

When the perturbation of the electric field is considered,
the instability can be related with the destruction of the black-hole structure.
It is known that the Cauchy (inner) horizon of the charged black hole
is unstable to form a singularity \cite{Gursel:1979zza}.
The perturbation introduced in this work may develop such an instability
in the RN black-hole type solution.

\section{Conclusions}
We investigated the gravitational field of static fluid plus electric field.
Both of the fluid and the electric field
are the sources of the gravitational field,
but the way to curve the spacetime is a bit different from each other.
By adopting the equation of state $p(r) = -\rho(r)/3$,
the fluid is responsible for the topology of the background space.
The spatial topology can be either closed ($S_3$) or open ($H_3$).
Such a nature of the spatial topology is not observed everywhere.
Instead, the signature of the background spatial topology appears
at some place of the spacetime.

Based on the background topology,
there exist various types of solutions in three classes
which we named as $S_3$-I, $S_3$-II, and $H_3$.
Interesting classes are $S_3$-I and $H_3$
although the class $S_3$-II has most varieties in solution.
The most interesting solutions are the black-hole solutions.
Due to the presence of the electric field,
the black-hole geometry mimics that of the Reisner-Norstr\"om spacetime.
This type of black hole exists in both $S_3$ and $H_3$ spaces.
(There exists also a Schwarzschild-type black hole in $S_3$-II.)
The central singularity inside the black hole of this type of solution
is due to the electric source as well as the fluid source.
There is a naked singularity in $S_3$-I at the antipodal point
which is not accessible except by the radial null rays.
The formation of this singularity is caused by the fluid.
The geodesics of the Reisner-Norstr\"om black-hole type solution
exhibit the oscillatory orbit in the infinite tower of the spacetime
encountered in the usual Reisner-Norstr\"om geometry.

All the solutions obtained in this paper are unconditionally unstable.
This is not surprising because the stability story is similar to
the fluid-only case in Ref.~\cite{Cho:2016kpf}.
The reason of the instability is that the static fluid becomes
unstable (time dependent) with small perturbations
and drives the background geometry to the Friedmann expansion.
In addition, there is an electric field for which it is well known
that the pure charged black-hole solution (Reisner-Norstr\"om geometry)
is unstable under perturbations.

The solutions investigated in this paper are
useful in studying the magnetic monopole in the closed/open space,
which is under investigation currently.
Usually, the outside geometry of the magnetic monopole is the same with that of
the charged black hole (Reisner-Norstr\"om geometry)
\cite{Gibbons:1990um,Cho:1975uz,Bais:1975gu,Yasskin:1975ag,Cordero:1976jc}.
Since we obtained the charged black-hole solution in $S_3$/$H_3$ with the aid of fluid,
it is very interesting to investigate the magnetic monopole in the presence of fluid.
It may give rise to insight about the monopole in the closed/open space.
The asymptotic geometry of this type of the gauge monopole is worth while to investigate
and will be very interesting to compare with the usual monopole geometry.
In addition, the removal of the singularity is also a very interesting issue.
For the usual case, the monopole field removes the singularity of the charged solution.
For this case, however, the formation of the singularity is caused
not only by the electric charge, but also by the fluid.
It is interesting to see if the monopole field can regularize the singular behavior of the fluid.

\acknowledgements
The author is grateful to Hyeong-Chan Kim and Gungwon Kang for useful discussions.
This work was supported by the grant from the National Research Foundation
funded by the Korean government, No. NRF-2017R1A2B4010738.


\begin{thebibliography}{99}

\bibitem{Ade:2015xua}
  P.~A.~R.~Ade {\it et al.} [Planck Collaboration],
  Astron.\ Astrophys.\  {\bf 594}, A13 (2016)
  doi:10.1051/0004-6361/201525830
  [arXiv:1502.01589 [astro-ph.CO]].


\bibitem{Ellis:2002we}
  G.~F.~R.~Ellis and R.~Maartens,
  Class.\ Quant.\ Grav.\  {\bf 21}, 223 (2004)
  doi:10.1088/0264-9381/21/1/015
  [gr-qc/0211082].

\bibitem{Ellis:2003qz}
  G.~F.~R.~Ellis, J.~Murugan and C.~G.~Tsagas,
  Class.\ Quant.\ Grav.\  {\bf 21}, no. 1, 233 (2004)
  doi:10.1088/0264-9381/21/1/016
  [gr-qc/0307112].

\bibitem{Labrana:2013oca}
  P.~Labrana,
  Phys.\ Rev.\ D {\bf 91}, no. 8, 083534 (2015)
  doi:10.1103/PhysRevD.91.083534
  [arXiv:1312.6877 [astro-ph.CO]].

\bibitem{Bucher:1994gb}
  M.~Bucher, A.~S.~Goldhaber and N.~Turok,
  Phys.\ Rev.\ D {\bf 52}, 3314 (1995)
  doi:10.1103/PhysRevD.52.3314
  [hep-ph/9411206].


\bibitem{White:2014aua}
  J.~White, Y.~l.~Zhang and M.~Sasaki,
  Phys.\ Rev.\ D {\bf 90}, no. 8, 083517 (2014)
  doi:10.1103/PhysRevD.90.083517
  [arXiv:1407.5816 [astro-ph.CO]].

\bibitem{Cho:2016kpf}
  I.~Cho and H.~C.~Kim,
  Phys.\ Rev.\ D {\bf 95}, no. 8, 084052 (2017)
  doi:10.1103/PhysRevD.95.084052
  [arXiv:1610.04087 [gr-qc]].

\bibitem{Bekenstein:1971ej}
  J.~D.~Bekenstein,
  Phys.\ Rev.\ D {\bf 4}, 2185 (1971).
  doi:10.1103/PhysRevD.4.2185

\bibitem{Sorkin:1981wd}
  R.~D.~Sorkin, R.~M.~Wald and Z.~J.~Zhang,
  Gen.\ Rel.\ Grav.\  {\bf 13}, 1127 (1981).
  doi:10.1007/BF00759862

\bibitem{Pesci:2006sb}
  A.~Pesci,
  Class.\ Quant.\ Grav.\  {\bf 24}, 2283 (2007)
  doi:10.1088/0264-9381/24/9/009
  [gr-qc/0611103].

\bibitem{Semiz:2008ny}
  I.~Semiz,
  Rev.\ Math.\ Phys.\  {\bf 23}, 865 (2011)
  doi:10.1142/S0129055X1100445X
  [arXiv:0810.0634 [gr-qc]].

\bibitem{Lake:2002bq}
  K.~Lake,
  Phys.\ Rev.\ D {\bf 67}, 104015 (2003)
  doi:10.1103/PhysRevD.67.104015
  [gr-qc/0209104].

\bibitem{Bronnikov:2008ia}
  K.~A.~Bronnikov and O.~B.~Zaslavskii,
  Phys.\ Rev.\ D {\bf 78}, 021501 (2008)
  doi:10.1103/PhysRevD.78.021501
  [arXiv:0801.0889 [gr-qc]].
  
\bibitem{Cho:2017nhx} 
  I.~Cho and H.~C.~Kim,
  arXiv:1703.01103 [gr-qc].

\bibitem{Misner:1964je} 
  C.~W.~Misner and D.~H.~Sharp,
  Phys.\ Rev.\  {\bf 136}, B571 (1964).
  doi:10.1103/PhysRev.136.B571
  
\bibitem{Gursel:1979zza} 
  Y.~Gursel, V.~D.~Sandberg, I.~D.~Novikov and A.~A.~Starobinsky,
  Phys.\ Rev.\ D {\bf 19}, 413 (1979).
  doi:10.1103/PhysRevD.19.413
  
\bibitem{Gibbons:1990um}
  G.~W.~Gibbons,
  Lect.\ Notes Phys.\  {\bf 383}, 110 (1991)
  doi:10.1007/3-540-54293-0$\_$24
  [arXiv:1109.3538 [gr-qc]].


\bibitem{Cho:1975uz}
  Y.~M.~Cho and P.~G.~O.~Freund,
  Phys.\ Rev.\ D {\bf 12}, 1588 (1975)
  Erratum: [Phys.\ Rev.\ D {\bf 13}, 531 (1976)].
  doi:10.1103/PhysRevD.13.531.2, 10.1103/PhysRevD.12.1588

\bibitem{Bais:1975gu}
  F.~A.~Bais and R.~J.~Russell,
  Phys.\ Rev.\ D {\bf 11}, 2692 (1975)
  Erratum: [Phys.\ Rev.\ D {\bf 12}, 3368 (1975)].
  doi:10.1103/PhysRevD.12.3368.2, 10.1103/PhysRevD.11.2692

\bibitem{Yasskin:1975ag}
  P.~B.~Yasskin,
  Phys.\ Rev.\ D {\bf 12}, 2212 (1975).
  doi:10.1103/PhysRevD.12.2212

\bibitem{Cordero:1976jc}
  P.~Cordero and C.~Teitelboim,
  Annals Phys.\  {\bf 100}, 607 (1976).
  doi:10.1016/0003-4916(76)90074-9

\end{thebibliography}
\end{document}